\documentclass[11pt, a4papper]{article}
\usepackage{amsmath,amsfonts,amssymb,amsthm,enumerate,graphicx,xcolor,url}
\usepackage{mathtools}
\usepackage{hyperref}
\usepackage{tikz}
\usepackage{booktabs}% nicer horizontal lines
\usepackage{caption}% fix vertical spacing of table captions
\usepackage{siunitx}% align numbers at the decimal point
\usepackage{tabularx}
\usepackage{graphicx}
\usepackage{adjustbox}
\usepackage{multirow}
\usepackage[pagewise]{lineno}
\newtheorem{theorem}{Theorem}[section]
\newtheorem{lemma}[theorem]{Lemma}

\usetikzlibrary{trees,arrows,positioning,fit,calc}
\tikzset{block/.style={draw, thick, text width=2cm, minimum height=0.75cm, align=center}
}

\theoremstyle{definition}
\newtheorem{definition}[theorem]{Definition}
\newtheorem{example}[theorem]{Example}

\theoremstyle{proposition}
\newtheorem{proposition}[theorem]{Proposition}
\theoremstyle{remark}
\newtheorem{remark}[theorem]{Remark}

\theoremstyle{corollary}
\newtheorem{corollary}[theorem]{Corollary}

\numberwithin{equation}{section}
\begin{document}

\title{Multisecret-sharing scheme with two-level security and its applications in Blockchain}
% A two level secure Blockchain based Secret sharing scheme
\author{ R. K. Sharma, R. Sarma, Neha Arora and Vidya Sagar}

\date{}

\maketitle

\vspace{-12mm}
\begin{center}

\noindent {\small Department of Mathematics,\\ Indian Institute of Technology Delhi,\\ Hauz Khas, New Delhi-110016, India$^{1}$.}

\end{center}

\footnotetext[1]{{\em E-mail addresses:} \url{rksharmaiitd@gmail.com} (Rajendra Kumar Sharma), \url{ritumoni407@gmail.com} (Ritumoni Sarma), \url{nehaarora1907@gmail.com} (Neha Arora), \url{vsagariitd@gmail.com} (Vidya Sagar).}

\medskip

\begin{center}
\date{\today}
\end{center}

\hrule

\begin{abstract}
%We investigate multisecret-sharing scheme based on Shamir's secret sharing scheme. In this scheme, we generalize Shamir's secret sharing scheme to two level security. Further, we generalize above scheme for multi-secret.

A $(t,m)$-threshold secret sharing and multisecret-sharing scheme based on Shamir's SSS are introduced with two-level security using a one-way function. Besides we give its application in smart contract-enabled consortium blockchain network. The proposed scheme is thoroughly examined in terms of security and efficiency. Privacy, security, integrity, and scalability are also analyzed while applying it to the blockchain network.
%A through analysis of our scheme on the basis of security and efficiency, and on the basis of privacy, security, and scalability when applied on Blockchain network has also been made.
\end{abstract}

\noindent {\small {\em MSC 2020\,:} Primary 94A62; Secondary 94A60.

\noindent {\em Keywords:} Secret sharing scheme, Multisecret-sharing scheme, One-way function, Hash Function, Blockchain, Smart Contract.}

\medskip

\section{INTRODUCTION}
Let $m, t\in \mathbb{N}$ with $1<t\leq m$. In secret sharing scheme, we divide the given data $\textbf{s}$, which is the \textit{secret}, into $m$ pieces, each of which is called as \textit{share}. One can obtain $\textbf{s}$ with the help of any $t$ or more shares but not with $t-1$ or less shares. Such scheme is called as a ($t, m$)-\textit{threshold secret sharing scheme}. Various cryptosystems which are based on single key have many shortcomings. For example, if the key is maliciously or accidentally disclosed to the public, or if the key's owner is found to be untrustworthy \cite{Raylin}, the entire system will be jeopardized, thus secret sharing schemes (SSS) are becoming essential nowadays; in fact, these are used heavily in electronic voting systems, cryptographic protocols, banking systems, etc.\par
In 1979, Blakley \cite{Blakley} and Shamir \cite{Shamir} introduced secret sharing scheme independently. To solve secret sharing problem, linear projective geometry was used in \cite{Blakley} whereas Lagrange interpolation polynomial was used in  \cite{Shamir}. After that, this topic got a lot of attention and researchers investigated various types of SSS, including On-line Schemes , Rational Schemes, Quantum Schemes, Chinese Remainder Schemes, Visual Schemes, Multiple Schemes, Proactive Schemes, Verifiable Schemes, Ideal Schemes, Linear Schemes, etc. in recent past.
%In \cite{Ebri_Yeun}, the authors used hyperplane over finite field instead of $\mathbb{R}$ to solve the secret sharing problem. 
Multisecret-sharing schemes (MSS) are well-known among the secret sharing scheme families. As the name suggests, in MSS \cite{Chien_Jan, Yang_Chang}, we have multiple secrets to be shared instead of a single secret.\par
 Blockchain is an open, decentralized, and distributed ledger that can record transactions among multiple parties in an efficient way. Each transaction is hashed (that is, digitally signed with cryptographically secured function) and then stored in blocks. Each block is linked with the previous block hash that makes it immutable. The concept of blockchain came into light in 2008, when a white paper \cite{Bitcoin} was published on virtual cryptocurrency Bitcoin by an anonymous person Satoshi Nakamoto. Later, it was made functional in 2009. %Later various other cryptocurrencies like Litecoin, Bytecoin, Peercoin, Emercoin, Ripple,e etc. were designed based on Blockchain Technology.
To enhance the privacy and security of blockchain, smart contracts \cite{smart contract} were used. It was first introduced by Nick Szabo in \cite{smart contract 1}.\par
In \cite{Raman_Varshney}% ref: Distributed Storage Meets Secret Sharing on the Blockchain
, the authors used Shamir's secret sharing scheme to distribute transaction data using private key encryption and distributed storage blockchain, keeping the data integrity intact. Similarly, in \cite{local sss}% ref: Efficient Local Secret Sharing for Distributed Blockchain Systems
, the authors proposed local secret sharing and applied it on distributed storage blockchain with the aim of reducing the storage and communication cost. A few applications of secret sharing schemes on the blockchain network have also been proposed in various sectors like healthcare \cite{Healthcare}, smart city architecture \cite{Smart city}, supply chain \cite{Supply chain}, etc. Their objective is to ensure the privacy and security of data from adversaries. \par
% ref: Blockchain Based Secret-Data Sharing Model for Personal Health Record System
% ref: Blockchain-empowered cloud architecture based on secret sharing for smart city
However, a secret sharing scheme has the limitation of dishonest dealers or participants, but none of them discussed about dishonest dealers and participants simultaneously. Also it is important to ensure that both the dealer's committee and participants (or miners) are honest. Since dealer plays the central role, it is assumed that the dealer must be honest. Also, in recent years, many blockchain based secret sharing schemes were introduced to outcome the limitation of dishonest participants. To protect the secret from attackers, many authors used the threshold $t/m$ less than $1/2$ \cite{BDLO15,VAE22,HJK95,MZW19, WWW02}. \par % ref: Storing and Retrieving Secrets on a Blockchain
In \cite{Benhamouda}% Ref: Can a public blockchain keep a secret?
 , the authors have discussed Dynamic Proactive Secret Sharing (DPSS) scheme, where dealers and participants keep on changing %and secret also keeps on changing
 and it is based on honest majority. Then, they have discussed Evolving-Committee Proactive Secret Sharing (ECPSS) scheme, which is a combination of DPSS and committee-selection protocol. They assume that either PoW or Cryptographic sortition can be done to choose the desired committee. It decreases the probability of corrupt members in the committee and restricts the adversary to know about the committee.
 % However self selection doesn't work well as there is a probability of corrupt members in committee or it will sometime allows adversary to know about the committee also.
Then, they have defined Target-Anonymous Channels, which keep the receivers (participants who receive shares of the secret) anonymous.\par
% In their technique, the blockchain maintains  a secret signature key and the holding committees use it to sign the blocks.
% They have also described various applications of Secret Sharing in Blockchain-specific contexts.

In this manuscript, first we define SSS and MSS based on Shamir's SSS with two-level security, where initially we check the honesty of participants. Further, only honest participants will get their share for the computation of the secret $\textbf{s}$. Then we apply our scheme on blockchain network. For this, we replace dealer with a team (or committee) of dealers, who need to prove their honesty using non-interactive zero-knowledge proofs before involving in the process of secret generation and distribution. Also, for the generation of a new block, a fresh committee will be formed, depending on the nodes involved in the transaction process. However, committee will have a predetermined minimum and maximum number of nodes and committee keeps on changing with the change in block. Moreover, $m$ participants will be chosen in an anonymous way. Once any $t$ out of them are able to retrieve the secret and validate the transactions, a new block will be formed and added to the chain.
In this case, if there are a few cheating participants and they try to find the secret by involving themselves in the secret recovery process, even then they will not be able to proceed to find the secret unless they retrieve the correct %hash
encrypted value of the secret, where encryption is done using a one-way function.\par
The rest of the manuscript is arranged as follows. Section \ref{section2} includes preliminaries. In Section \ref{section3}, we propose our scheme. Application of the scheme on blockchain network is discussed in section \ref{section4}. In Section \ref{section5}, we analyze our scheme on the basis of its efficiency and security. Also, we examine the privacy, security, integrity, and scalability of the scheme while applying it on blockchain in this section. Section \ref{section6} concludes the manuscript.

\section{Definitions and Preliminaries}\label{section2}

% Secret sharing
\begin{definition}
\textit{Secret sharing scheme} is a way in which one (called \textit{dealer}) distributes the secret to multiple people (called \textit{participants}) in such a way that they can collectively recover the secret but individually they can't.\\
Let the secret be distributed to $m$ participants. If the secret can only be recovered by any $t$ or more participants then $t$ is said to be the \textit{threshold} of the scheme, where $1<t \leq m$.\\
A $(t, m)$-\textit{threshold secret scheme} is a scheme with threshold $t$ and $m$ participants.
\end{definition}
\noindent\textbf{Shamir's Secret sharing scheme} \cite{Shamir}:
In this scheme, Shamir has taken two entities: a dealer and a set of participants. Dealer is the one who knows the secret \textbf{s} and distributes their shares (pieces of the secret) to all participants in such a way that it follows the basic properties of SSS and its threshold. For this, he has constructed a polynomial $f(x)$ of degree $t-1$ such that constant term of the polynomial is $\textbf{s}$ and the remaining coefficients are randomly generated.
\begin{itemize}
\item Secret generation: Dealer chooses random variables, say $r_1, r_2, \dots ,r_{t-1}$ and generates the polynomial $f(x) = \textbf{s} + r_1 x + r_2 x^2 + \dots + r_{t-1} x^{t-1}$ and then enumerates the participants and computes $f(1), f(2),\dots, f(m)$.
\item Secret distribution: Distribute $(i, f(i))$ to the $i^{th}$ participant, $ 1\leq i \leq m$.
\item Secret recovery: Any $t$ or more participants come together to combine their shares $(i,f(i))$ to compute the polynomial $f(x)$ with the help of Lagrange interpolation polynomial.
\end{itemize}

\begin{definition}
A $(t, m)$-threshold secret sharing scheme is said to be \textit{perfect} if any $\tilde{k}$ (where $\tilde{k} < t$) participants cannot recover the secret.
\end{definition}

\begin{definition}
\textit{Hash function} is a function which maps bit string of arbitrary length to a bit string of fixed length in random manner.
%\begin{equation*}
% H : \mathbb{Z}_2 ^* \rightarrow \mathbb{Z}_2 ^l   
 %\end{equation*}
 \\
\textit{Cryptographically secure hash function} is a hash function which satisfies the following properties: one-wayness, collision resistant, target collision resistant, non-malleability and pseudo-randomness.
\end{definition}
\begin{remark}
	Throughout this paper, by a hash function, we mean a Cryptographically secure hash function.
\end{remark}

\begin{definition}
A function $T: A \rightarrow B$ is said to be \textit{one-way function} if for any $b \in B$, it is computationally hard to find some $a \in A$ such that $T(a) = b$ in polynomial time.
\end{definition}

\begin{definition}
\textit{Distributed ledger} is a type of database which records the data or information in such a way that it will be shared and replicated in its most updated form to all the members available on the decentralized network. Participants in the network agree on the consensus and update the database timely using cryptographic signature, which makes it auditable for the remaining members.
\end{definition}
%%%%%%%%

%%%%%%%%%%%%%%

\begin{definition} \cite{block}
A \textit{block} is a container that contains a series of transactions. A block  is divided into two components: block header and transaction counter.
\\ Block header contains the following.
\begin{itemize}
\item Block version: It indicates the validation rules for the block.
\item Merkle tree root hash: The aggregate of hash value of all transactions.
\item Timestamp: Current time in seconds/minutes since the starting of the network.
\item nBits: It is related to the difficulty level for the computation of new hash and its size in bits.
\item Nounce: A variable which keeps on increasing with every hash calculation and PoW done.
\item Parent block hash: Hash value of the previous block.
\end{itemize}
Transaction counter contains the transactions. Maximum number of transactions that a block can contain depends on the size of the block, the size of each transaction, and the total number of transactions occurred in a fixed time interval.

\end{definition}
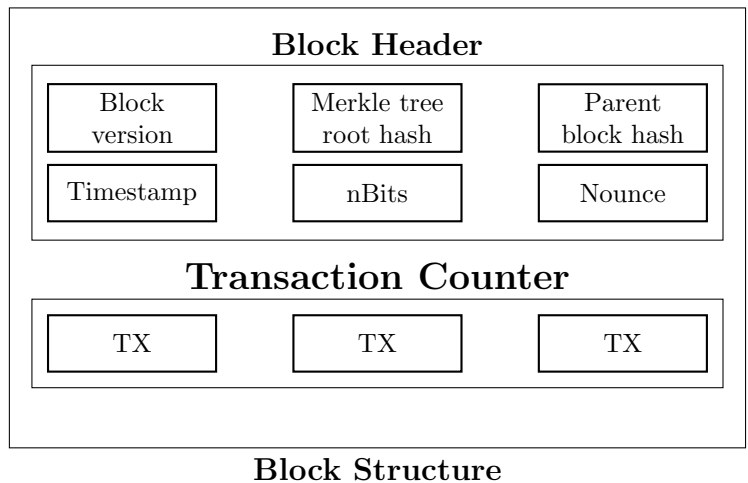
\begin{center}
	\begin{small}
		\begin{tikzpicture}
			\begin{scope}				
				\node[block] (a) {Block version};
				\node[block,right=of a] (b) {{Merkle tree root hash}};
				\node[block,right=of b] (c) {{Parent block hash}};
				%\node[block,right=of c] (d) {nBits};
				\node[block] (d) at ([yshift=-1cm]$(b)!1!(a)$) {Timestamp};
				\node[block,right=of d] (n) {nBits};
				\node[block,right=of n] (v) {Nounce};
				\node[block] (e) at ([yshift=-3cm]$(b)!1!(a)$) {TX};
				\node[block,right=of e] (g) {TX};
				\node[block,right=of g] (h) {TX};
				%\node[block,right=of h] (i) {TX};
				%\node[block] (j) at ([yshift=-8cm]$(b)!1!(a)$) {T5};
				%\node[block,right=of j] (k) {T6};
				%\node[block,right=of k] (l) {T7};
				%\node[block,right=of l] (m) {T8};
				\node[draw,inner xsep=5mm,inner ysep=10mm,fit=(a)(b) (c) (d) (n) (v) (g) (h),label={270:\textbf{{\large Block Structure}}}]{};
				\node[draw,inner xsep=2mm,inner ysep=2mm,fit=(a) (b) (c) (d) (n) (v),label={90: \textbf{{\large Block Header}}}]{\vspace{2 cm}};
				\node[draw,inner xsep=2mm,inner ysep=2mm,fit=(e) (g) (h),label={90: \textbf{{\Large Transaction Counter}}}]{};

			\end{scope}
		\end{tikzpicture}
	\end{small}
\end{center}

\begin{definition}
A \textit{smart contract} is a digital contract which enforces the members to follow the procedure and instructions mentioned in the contract. It ensures smooth functioning of the system and verification of the transactions without any interruption from the outside network. They are immutable and distributed. It also provides security to the network.
\end{definition}

\begin{definition}
A \textit{zero-knowledge proof} is a protocol by which a prover can prove the knowledge of certain information (say, $x$) without revealing it and the verifier can verify it without getting any information about $x$. A zero knowledge proof must posses three properties: completeness, soundness, and zero-knowledge.
\end{definition}

\section{Proposed Schemes}\label{section3}
%%%%%%%%%%
\begin{center}
	\begin{small}
		\begin{tikzpicture}
			\begin{scope}				
				\node[block] (a) {Dealer \\(or System)};
				%\node[block,right=of a] (b) {{System}};
				\node[block,right=of a] (b) {{Participants}};
				\node[block,right=of b] (c) {{System}};
				\node[block,right=of c] (d) {{Participants}};
				\draw[->] (a)-- (b); 
				%\draw[->] (a)-- node{s} (b); 
				\draw[->] (b)-- (c);  
				\draw[->] (c)-- (d);  
				%\draw[->] (d)-- (e);   
				\node[draw,inner xsep=2mm,inner ysep=10mm,fit=(a)(b) (c) (d),label={270:\textbf{{\large Communication Channel}}}]{};
				
			\end{scope}
		\end{tikzpicture}
	\end{small}
\end{center}

%%%%%%%%%%%%%%%%%
In previous secret sharing schemes, without testing the honesty of the participant, it is assumed that majority of them are honest and thus they become eligible to compute the secret. However the hypothesis of honest participants may fail. Here we have defined two steps secret sharing scheme where the first step is only for verifying if the active participants are honest or not. In this step, instead of sending the shares for the computation of the secret $\textbf{s}$, dealer(or system) shares the information (on similar lines as we have done in section \ref{section3.1} and \ref{section3.2} or using any available SSS) for the computation of $H(\textbf{s})$, where $H$ is a one-way function.
Once $H(\textbf{s})$ is computed correctly, the system shares the information, for the computation of the secret $\textbf{s}$, only to those participants, who participated in the computation of $H(\textbf{s})$ and then they can collectively compute the secret $\textbf{s}$.

We have proposed the secret sharing scheme by applying the above procedure on Shamir's SSS for the single secret $\textbf{s}$ and for multi secret $\textbf{s} = (s_1, s_2, \dots, s_m)$. 

%In this paper, we have proposed the secret sharing scheme with two level security.

\subsection{Generalization of Shamir's Secret Sharing Scheme with two-level security}\label{section3.1}

\subsubsection{Set up Phase:}\label{section3.1.1}
In this scheme, we have $\{P_1, P_2, P_3, \dots , P_m\}$ as $m$ participants and $(t,m)$ is the threshold. Dealer D (can be replaced by the system S) chooses $a_i \in \mathbb{F}_p^{\ast}$, where $p$ is a large prime such that $m << p$, to be the public key of $P_i$ respectively such that $a_i \neq a_j$ for $i \neq j$. Let $\textbf{s} \in \mathbb{F}_p $ be the secret.
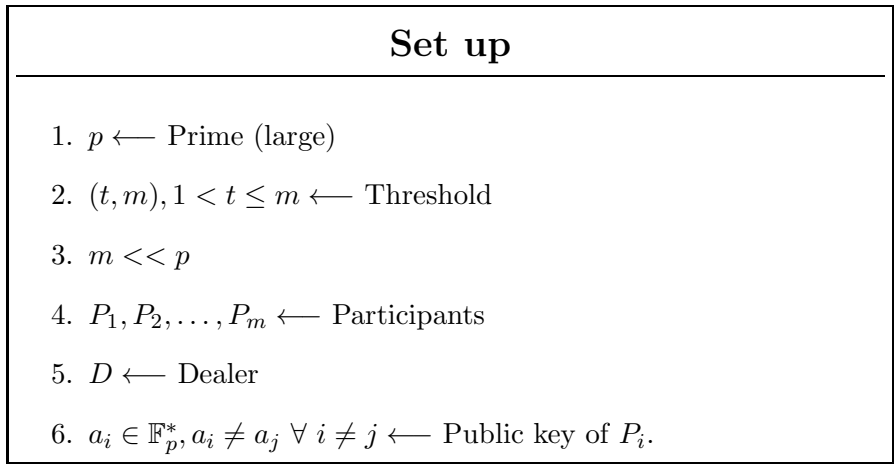
\begin{center}
	\fbox{\begin{minipage}{30em}
			\begin{center}
				\vspace{0.2 cm}
				{\Large \textbf{Set up}}\\
				\vspace{-0.2 cm}
				\hrulefill
			\end{center}
			
			\begin{enumerate}
				\item $p \longleftarrow $ Prime (large)
				\item $(t,m) , 1 < t \leq m \longleftarrow $ Threshold
				\item   $m < < p $
				\item $P_1, P_2, \dots , P_m \longleftarrow $ Participants
				\item $D \longleftarrow $ Dealer
				\item $a_i \in \mathbb{F}_p^{\ast} , a_i \neq a_j \  \forall \  i \neq j \longleftarrow $ Public key of $P_i$.
			\end{enumerate}
	\end{minipage}}
\end{center}
\subsubsection{Computing and distributing shares:}
Dealer chooses a one-way function $H$, and random elements $r_1, r_2, \dots , r_{t-1} \in \mathbb{F}_p $, and then computes $H(r_1), H(r_2), \dots  , H(r_{t-1})$. He then generates the polynomials $f(x)$ and $h(x)$ as follows:
\begin{equation}
	f(x) \ = \ \textbf{s} + r_1 x + r_2 x^2 + \dots + r_{t-1} x^{t-1}  \ \ \ \ \ \ \ \ \ \ \ \ \ \ \ \ \ \ \ \ 
\end{equation}
\begin{equation}
	h(x) \ = \ H(\textbf{s}) + H(r_1) x + H(r_2) x^2 + \dots + H(r_{t-1}) x^{t-1}
\end{equation}
Dealer then computes $(h(a_i), f(a_i))$ and initially shares $h(a_i)$ to the participant $P_i$ ($1 \leq i \leq m$).

\begin{center}
	\fbox{\begin{minipage}{30em}
			\begin{center}
				\vspace{0.2 cm}
				\textbf{Dealer (for single secret)}\\
				\vspace{-0.2 cm}
				\hrulefill
			\end{center} 
			\begin{enumerate}
				\item $H \longleftarrow$ One-way function
				\item $r_1, r_2, \dots , r_{t-1} \in \mathbb{F}_p \longleftarrow $ Random elements
				\item $\textbf{s} \longleftarrow $ Secret
				\item Compute $H(r_1), H(r_2), \dots , H(r_{t-1}) $
				\item Generate $f(x)$ and $h(x)$ by: \\
				$f(x) \ = \ \textbf{s} + r_1 x + r_2 x^2 + \dots + r_{t-1} x^{t-1}$ \\
				$h(x) \ = \ H(\textbf{s}) + H(r_1) x + H(r_2) x^2 + \dots + H(r_{t-1}) x^{t-1}$
			\end{enumerate}		
			\hrulefill
			\vspace{-0.3 cm}
			\begin{center}
				Output : $(a_i, f(a_i), h(a_i)) ; 1 \leq i \leq m  \longrightarrow $ System 
			\end{center}
			
	\end{minipage}}
\end{center}

\subsubsection{Recovering the secret:}
Any $t$ or more participants, upon receiving $h(a_i)$ corresponding to their public key $a_i$, can come forward to compute the polynomial $h(x)$ by using Shamir's secret sharing scheme and then share its constant term $H(\textbf{s})$ to the system.
\\
System verifies $H(\textbf{s})$ and after confirming the honesty of minimum of $t$-participants, it reveals $f(a_i)$ to only those $t$-participants $P_i$, who take part in computation of $h(x)$ and passes the honesty test.  Then, they can finally recover the secret $\textbf{s}$.

\begin{center}
	\fbox{\begin{minipage}{30em}
			\begin{center}
				\textbf{Participants} : $t \longleftarrow$ minimum number of participants required to recover the secret\\
				\vspace{-0.2 cm}
				\hrulefill
			\end{center} 
			\begin{enumerate}
				\item At least $t$ participants interact with the system to find the secret.
				\item Input: $(a_i, h(a_i))$  by at least $t$ participants
				\item Output: $h(x) \longrightarrow $ System
				\item  System verifiers $h(x)$ % and if correct, send $f(a_i)$ to participating members
				 \\  if:
				\\ less than $t$ participants are honest
				\\ STOP
				\\  else: 
				\\  send $f(a_i)$ to participating members
				\item Input: $(a_i, f(a_i))$  %by participating members
				\item Output: $f(x)$
				\item Compute the secret $\textbf{s}$.
			\end{enumerate}
	\end{minipage}}
\end{center}

% Ref: Several Generalization of Shamir's Secret Sharing Scheme

%%%%%%%%%%%%%%%%%%%%%%%%%%%%%%%%%%
%%%%%%%%%%%%%%%%%%%%%%%%%%%%%%%%%%
%%%%%%%%%%%%%%%%%%%%%%%%%%%%%%%%%%
%%%%%%%%%%%%%%%%%%%%%%%%%%%%%%%%%%

\subsection{Generalization of Shamir's Secret Sharing Scheme for multi-secret with two-level security}\label{section3.2}

\subsubsection{Set up Phase:}
In this scheme, we have assumed the same set up as we have done in \ref{section3.1.1} for single secret. Let $\textbf{s} = (s_1, s_2, \ . \ . \ . \ s_m)$ be the secret such that $s_i \in \mathbb{F}_p $. Also, the first $k \ (1 < k \leq t) $ bits of $\textbf{s}$ are message bits and remaining $m-k$ bits are parity bits with $t \leq m-1$.

\subsubsection{Computing and distributing shares:}
Dealer chooses a one-way function $H$. 
He computes  $\tilde{s}  \ = \ \displaystyle\sum_{i = 1}^m s_i$ and makes it public. He then computes  $\alpha_i$, and $H(\alpha_i)$ for each $i \in \{1,2, \dots , t\}$, where
\begin{equation*}
	\alpha_i  \ = \ \displaystyle\sum\limits_{\substack{j= 1  \\  j \neq i}}^m s_j.
\end{equation*}
He then generates the polynomials $f(x)$ and $h(x)$ as follows:
\begin{equation}
	f(x)  =  \alpha_1 + \alpha_2 x + \alpha_3 x^2 + \dots + \alpha_t x^{t-1}, \ \ \ \ \ \ \ \ \ \ \ \ \ \ \ \ \ \ \ \ 
\end{equation}
\begin{equation}
	h(x) =  H(\alpha_1) + H(\alpha_2) x + H(\alpha_3) x^2 + \dots + H(\alpha_t) x^{t-1}.
\end{equation}
Dealer then computes $(h(a_i), f(a_i))$, where $a_i$ is the public key of $i^{th}$ participant $P_i$ and initially shares $h(a_i)$ to the participant $P_i$ ($1 \leq i \leq m$).

\begin{center}
	\fbox{\begin{minipage}{30em}
			\begin{center}
				\vspace{0.2 cm}
				\textbf{Dealer (for multi secret)}\\
				\vspace{-0.2 cm}
				\hrulefill
			\end{center} 
			\begin{enumerate}
				\item $H \longleftarrow$ One-way function
				\item $\textbf{s} = (s_1, s_2, \dots, s_m) \longleftarrow $ Secret such that the first $k$ bits $(1 < k \leq t)$ are message bits and the remaining $n-k$ bits are parity bits.
				\item Compute $\tilde{s} = \ \displaystyle\sum_{i = 1}^m s_i \ \longleftarrow $ Public to all 
				\item Compute $ \alpha_i =   \ \displaystyle\sum\limits_{\substack{j= 1  \\  j \neq i}}^m s_j $ where $ 1 \leq i \leq t$
				\item Compute $H(\alpha_1), H(\alpha_2), \dots, H(\alpha_{t}) $
				\item Generate $f(x)$ and $h(x)$ by: \\
				$f(x) \ = \  \ \alpha_1 + \alpha_2 x + \alpha_3 x^2 + \dots + \alpha_t x^{t-1} $ \\
				$h(x) \ = \ \ H(\alpha_1) + H(\alpha_2) x + H(\alpha_3) x^2 + \dots + H(\alpha_t) x^{t-1}$
			\end{enumerate}
			\hrulefill
			\vspace{-.3 cm}
			\begin{center}
				Output : $ (a_i, f(a_i), h(a_i)) \  \forall \ 1 \leq i \leq m  \longrightarrow $ system 
			\end{center}
			
	\end{minipage}}
\end{center}

\subsubsection{Recovering the secret:}
Any $t$ or more participants, upon receiving $h(a_i)$ corresponding to their public key $a_i$, can come forward to compute the polynomial $h(x)$ by using Shamir's secret sharing scheme and then share it with the system.
\\
System verifies $h(x)$ and after confirming the honesty of minimum of $t$-participants, it reveals $f(a_i)$ to only those $t$-participants, who take part in the computation of $h(x)$ and passes the honesty test. Then, they can recover the polynomial $f(x)$.
\\
Once, $f(x)$ is recovered, participants can compute $s_i \ = \ \tilde{s} - \alpha_i \ \ \forall \ 1 \leq i \leq t $.

\iffalse \begin{center}
	\fbox{\begin{minipage}{30em}
			\begin{center}
				\textbf{Participants} : $t \longleftarrow$ minimum number of participants required to recover the secret\\
				\vspace{-0.2 cm}
				\hrulefill
			\end{center} 
			\begin{enumerate}
				\item $t$ participants interact with the system to find the secret.
				\item Input: $(a_i, h(a_i))$  by at least $t$ participants
				\item Output: $h(x) \longrightarrow $ System
				\item  System verifiers $h(x)$ and if correct, send $f(a_i)$ to participating members
				\item Input: $(a_i, f(a_i))$  by participating members
				\item Output: $f(x)$
				\item Compute the secret $\textbf{s}$.
			\end{enumerate}
	\end{minipage}}
\end{center} \fi

\begin{example}	
	Suppose $p=199$. Let $(5,11)$ be the threshold and $P_1, P_2, P_3, \dots .P_{11} $ be the $11$ participants and $(a_1, a_2, \dots , a_{11}) \ = \ (7,5,4,3,2,9,6,8,11,10,12)$, where $a_i$ is the public key of $P_i$.
	\\
	Let $s = (7,9,2,3,7,5,4,9,3,21,27)$ be the secret, where only the first $5$ bits are message bits and the remaining $6$ bits are parity bits.
	\\
	Suppose $H: \mathbb{F}_{199}\longrightarrow \mathbb{F}_{199}$ defined by $H(n)=g^n$, where $g\in \mathbb{F}_{199}^{\ast}$ is the one-way function. Take $g=3$.
	\\
	Then $\tilde{s} = \displaystyle\sum_{i = 1}^{11} s_i  = 97 $ and it is made public.
	\\
	Further, $\alpha_1 = 90 ,  \alpha_2 = 88 ,  \alpha_3 = 95 , \alpha_4 = 94  , \alpha_5 = 90$ and $H(\alpha_1) = 188, H(\alpha_2) = 43 , H(\alpha_3) = 113 , H(\alpha_4) = 104 , H(\alpha_5) = 188 $ and therefore
	\begin{equation*}
		\begin{split}
			f(x) =  90 + 88x + 95 x^2 + 94 x^3 + 90 x^4
		\end{split}
	\end{equation*}
	and
	\begin{equation*}
		\begin{split}
			h(x) = 188 + 43x + 113 x^2 + 104 x^3 + 188 x^4.
		\end{split}
	\end{equation*}
	Dealer then computes $f(a_i)$ and $h(a_i)$ for each $i$ and share it with the system, which is displayed in the following table.	
	\begin{center}
		\begin{adjustbox}{width=\textwidth}
		\begin{tabular}{|c||c|c|c|c|c|c|c|c|c|c|c|}
			\hline
			$i$ & $1 $ & $ 2$ & $ 3$ & $5 $  & $5 $ & $6 $ & $ 7$ & $8 $ & $ 9$ & $10 $ & $11 $ \\ \hline
			$a_i$ & $ 7$ & $5 $ & $4 $ & $ 3$  & $2 $ & $9 $ & $ 6$ & $8 $ & $ 11$ & $10 $ & $12 $  \\ \hline
			$f(a_i) (\textnormal{mod 199}) $ & $ 167$ & $61$ & $173$ & $92$  & $52$ & $147$ & $90$ & $170 $ & $70$ & $117 $ & $ 116$  \\ \hline
			$h(a_i)(\textnormal{mod 199})$   & $163 $ & $0 $ & $38 $ & $67 $  & $ 188$ & $ 40$ & $185 $ & $36 $ & $ 65$ & $147 $ & $34 $  \\ \hline
		\end{tabular}
	\end{adjustbox}
	\end{center}	
	System initially shares $h(a_i) $ to $P_i$ for each $i$ and any $5$ or more participants can decrypt the polynomial $h(x)$. Without loss of generality, assume $P_1, P_2, P_3, P_4, P_5$ come together to recover the polynomial $h(x)$, then they have the following.
	\begin{center}
		\begin{tabular}{|c||c|c|c|c|c|}
			\hline
			$i$ & $1 $ & $ 2$ & $ 3$ & $4 $ & $5$ \\ \hline
			$a_i$ & $ 7$ & $5 $ & $4 $ & $ 3$  & $2 $ \\ \hline
			$h(a_i)$   & $163 $ & $0 $ & $38 $ & $67 $  & $ 188$  \\ \hline
		\end{tabular}
	\end{center}
	Since $h(x)$ is a polynomial of degree $4$ and the active participants have its value at $5$ distinct points, they use Lagrange Interpolation to compute the polynomial $h(x)$.
	% $$h(x) =  188+ 43x + 113 x^2 + 104 x^3 + 188 x^4.$$
	Participants will now share it with the system and after verifying, the system will share the corresponding values of the polynomial $f(x)$ to respective the participants as follows.
	\begin{center}
		\begin{tabular}{|c||c|c|c|c|c|}
			\hline
			$i$ & $1 $ & $ 2$ & $ 3$ & $4 $ & $5$ \\ \hline
			$a_i$ & $ 7$ & $5 $ & $4 $ & $ 3$  & $2 $ \\ \hline
			$f(a_i) $ & $ 167$ & $61$ & $173$ & $92$  & $52$  \\ \hline
		\end{tabular}
	\end{center}
	Participants will now compute the polynomial $f(x)$ by the same method.
	% $$ f(x) \ = \  90 + 88x + 95 x^2 + 94 x^3 + 90 x^4 $$
	Then, they compute $s_j \ = \ \tilde{s} - \alpha_j \ = \ 97 - \alpha_j \ \ \ \forall \ 1 \leq j \leq 5 $.
	
	\begin{center}
		\begin{tabular}{|c||c|c|c|c|c|}
			\hline
			$j$ & $1 $ & $ 2$ & $ 3$ & $4 $ & $5$ \\ \hline
			$\alpha_j$ & $ 90 $ & $ 88 $ & $ 95 $ & $ 94 $  & $ 90 $ \\ \hline
			$ s_j $ & $ 7 $ & $ 9 $ & $ 2 $ & $ 3 $  & $ 7 $   \\ \hline
		\end{tabular}
	\end{center}
	% Lagrange interpolation or by forming the system of equations.
\end{example}

\begin{example}
	
	Suppose $p = 113$. Let $(9,17)$ be the threshold and $P_1, P_2, P_3, \dots .P_{17}$ be the $17$ participants and $a_i \ = \ i $ be the public key of  $P_i$.
	\\
	Let $s = (3,5,7,9,11,3,5,6,2,1,7,8,6,2,5,1,4)$ be the secret, where only the first $6$ bits are message bits and remaining $11$ bits are parity bits.
	\\
	Note that message bits is less than the threshold.
	\\
	Suppose $H : \mathbb{F}_{113} \longrightarrow \mathbb{F}_{113} $ defined by $H(n) = n^2 \ (mod \ 113) $ is the one-way function.
	%Assume $H : \mathbb{F}_{13^3} \longrightarrow \mathbb{F}_{13^3} $ defined by $H(0)=0$ and $H(g^i)=i\ ( mod \ 7)$, where $g$ is the generator of the cyclic group $\mathbb{F}^{\ast} _{7^3} $.
	\\
	Then $\tilde{s} \ = \ \displaystyle\sum_{i = 1}^{17} s_i  = 85 \ (mod \ 113)  $ and it is made public.
	\\
	Further, $ \alpha_1 = 82  ,   \alpha_2 = 80  ,   \alpha_3 = 78  ,  \alpha_4 = 76  ,  \alpha_5 = 74  ,  \alpha_6 = 82  ,   \alpha_7 = 80  ,  \alpha_8 = 79  ,  \alpha_9 = 83 $ and $ H(\alpha_1) = 57   ,   H(\alpha_2) = 72  ,   H(\alpha_3) = 95  ,  H(\alpha_4) = 13  ,  H(\alpha_5) = 52  ,   H(\alpha_6) = 57  ,   H(\alpha_7) = 72  ,  H(\alpha_8) =  26  ,  H(\alpha_9) = 109 $ and therefore

	$$f(x) = 82 + 80x + 78x^2 + 76x^3 +74 x^4 + 82x^5 +80x^6+79x^7+83x^8 $$
	and
	\begin{equation*}
		\begin{split}
			h(x) =   57 + 72 x + 95 x^2 + 13 x^3 + 52 x^4 + 57 x^5 + 72 x^6 + 26 x^7 + 109 x^8 .
		\end{split}
	\end{equation*}
	Dealer then computes $f(a_i)$ and $h(a_i)$ for each $i$ and share it with the system, which is displayed in the following table.		
    \begin{center}
		\begin{tabular}{|c||c|c|c|c|c|c|c|c|c|c|c|c|c|c|c|c|c|}
			\hline
			$a_i$ & $1 $ & $ 2$ & $ 3$ & $4 $ & $5$ & $6$ & $7$ & $8$ & $9$ & $10$ & $11$ & $12$ & $13$ & $14$ & $15$ & $16$ & $17$ \\ \hline
			$f(a_i) $ & $ 36$ & $92$ & $92$ & $63$  & $16$ & $9$ & $28$ & $96$ & $68$  & $104 $ & $52 $ & $106 $ & $84 $ & $47 $ & $ 73$ & $3 $ & $ 41$ \\ \hline
			$h(a_i) $ & $101 $ & $83 $ & $44 $ & $108 $  & $ 1$ & $65 $ & $ 66$ & $89 $ & $100 $  & $ 37$ & $ 3$ & $105 $ & $19 $ & $27 $ & $45 $ & $ 25$ & $0 $ \\ \hline
		\end{tabular} 
	\end{center}	
	System initially shares $h(a_i) $ to $P_i$ for each $i$ and any $9$ or more participants can decrypt the polynomial $h(x)$. Without loss of generality, assume $P_1, P_2, \dots , P_9$ come together to recover the polynomial $h(x)$, then they have the following.	
    \begin{center}
		\begin{tabular}{|c||c|c|c|c|c|c|c|c|c|}
			\hline
			$a_i$ & $1 $ & $ 2$ & $ 3$ & $4 $ & $5$ & $6$ & $7$ & $8$ & $9$ \\ \hline
			$h(a_i) $ & $101 $ & $83 $ & $44 $ & $108 $  & $ 1$ & $65 $ & $ 66$ & $89 $ & $100 $ \\ \hline
		\end{tabular} 
	\end{center}	
	Since $h(x)$ is a polynomial of degree $8$ and the active participants have its value at $9$ distinct points, they use Lagrange Interpolation to compute the polynomial $h(x)$.
	Participants will now share it with the system and after verifying, the system will share the corresponding values of the polynomial $f(x)$ to the respective participants as follows.	
    \begin{center}
		\begin{tabular}{|c||c|c|c|c|c|c|c|c|c|}
			\hline
			$a_i$ & $1 $ & $ 2$ & $ 3$ & $4 $ & $5$ & $6$ & $7$ & $8$ & $9$ \\ \hline
			$f(a_i) $ & $ 36$ & $92$ & $92$ & $63$  & $16$ & $9$ & $28$ & $96$ & $68$ \\ \hline
		\end{tabular} 
	\end{center}	
	Participants will now compute the polynomial $f(x)$ by the same method.
	Then, they compute $s_j \ = \ \tilde{s} - \alpha_j \ = \ 85 - \alpha_j \ \ \ \forall \ 1 \leq j \leq 9 $.	
	\begin{center}
		\begin{tabular}{|c||c|c|c|c|c|c|c|c|c|}
			\hline
			$j$ & $1 $ & $ 2$ & $ 3$ & $4 $ & $5$ &$6$&$7$&$8$&$9$ \\ \hline
			$\alpha_j$ & $82$ & $ 80 $ & $ 78$ & $ 76 $  & $ 74 $ & $ 82 $ & $ 80 $ & $ 79 $& $ 83 $ \\ \hline
			$ s_j $ & $ 3 $ & $ 5 $ & $ 7 $ & $ 9 $  & $ 11 $ & $ 3 $ & $ 5 $ & $ 6 $ & $ 2 $  \\ \hline
		\end{tabular}
	\end{center}
	Since, only first $6$ bits are message bits, thus required message is $(3,5,7,9,11,3)$.
\end{example}

%%%%%%%%%%%%%%%%%%%%%%%%%%%%%%%%%%%%%%%
%%%%%%%%%%%%%%%%%%%%%%%%%%%%%%%%%%%%%%%
%%%%%%%%%%%%%%%%%%%%%%%%%%%%%%%%%%%%%%%
%%%%%%%%%%%%%%%%%%%%%%%%%%%%%%%%%%%%%%%

\section{Multisecret-sharing scheme on a Blockchain Network}\label{section4}
\subsection{Blockchain Architecture}
Blockchain is a chain of virtual blocks in which each block contains certain information along with its hash and the hash of the previous block. In this subsection, we will demonstrate how the blocks of blockchain are formed with the help of the proposed scheme. We impose a few assumptions and then define the structure of a blockchain network to efficiently apply this scheme as follows.

\begin{itemize}
	\item Type of blockchain:
	We assume that our platform is smart contract-enabled consortium blockchain network with limited number of members, referred to as \textit{nodes}. Each member is bound to follow the procedure written in smart contract and any kind of violation will lead to heavy penalty or cancellation of their participation. We refer any kind of exchange as \textit{transaction}.
	
	\item Structure of block:
	Each block is divided into two sections: Block Header and Transaction Counter. Block Header further contains block version, Merkle tree root hash, timestamp, nBits, nounce (secret), and parent block hash. Transaction Counter stores the transactions.
	
	\item Generation of a block:
	We assume that a new block is generated after a finite predetermined time once the secret is recovered corresponding to all the transactions done in a fixed time interval. 
	
	\item Channel:
	We have assumed our scheme as Evolving-Committee Proactive Secret Sharing Scheme and Channel as Target-Anonymous Channel, discussed in \cite{Benhamouda}.
	
	\item Dealer:
	There will be a team (or committee) of dealers, which is freshly formed for the secret generation and validation of each new block. It will be chosen on the bases of transactions occurred and PoW done, where nodes need to prove the validity of their transaction, using non-interactive zero-knowledge proofs (i.e. without revealing any information about the transactions) and then generate the shares of the secret to be distributed.
	
	\item Secret generation and distribution:
	To generate the secret $s_{B_i}$ for $i$-th block, dealers require the number of transactions in that particular time interval, the number of people involved in the transactions, transaction id's, and the total amount debited and credited. The secret will be distributed among random active nodes. 
	
	\item Participants:
	Since, our scheme follows Target-Anonymous Channel, the participants (who receive the secret share) are anonymous. Secret share will be distributed to few active nodes (miners) anonymously and they required to collectively participate, compute the secret, and verify and validate the transactions (called as \textit{mining process}). If participants were not able to conclude (that the transaction is valid or not) within the predetermined time interval, it would be considered as validated and automatically be added to the block and no further questioning will be allowed.	
	
	% no. of transactions in that particular time interval; no. of people involved in the transactions; transaction ids; amount debited and credited, verify if net is zero (debit - credit = 0);
	
	\item Secret recovery:
	 A threshold of $50\%$ is required to set, that is, at least $50\%$ active nodes need to find the secret. Once they recover the secret, they need to verify and validate the transactions.
	
	\item Formation of block:
	Once, the transactions are validated, a new block will be formed and added to the longest available chain, containing all the validated transactions stored in it.	
	
\end{itemize}

\begin{example}
We will show it with an example.
\begin{enumerate}
\item Assume that we have $100$ participants in our blockchain network and each member is given the identity $U_i$ , $1 \leq i \leq 100$.
\item A new block is generated after every $\tau_{\circ}$ minutes, and the active nodes (miners) will be given $\tau_{1}$ minutes to recover the secret that validates the transactions.
\item Let $T_{i,j}$ be the transaction ID of the $j^{th}$ transaction for the $i^{th}$ block.
%\item Let $T_{i,j}$ be the transaction id where $i$ represents the block number and $j$ represents the transaction number of $i$-th block.
\item Let $T_i$ be the concatenation of all the transaction IDs of the $i^{th}$ block.
\item Suppose there are $20$ transactions that has taken place during the $k^{th}$ interval $[(k-1)\tau_{\circ}, k\tau_{\circ} \ ]$ and identities involved in these transactions are $U_1,U_2, U_3, \dots , U_{14}$.
\item Then $U_1,U_2, U_3, \dots , U_{14}$ form the committee of dealers and they need to validate the transactions using zero-knowledge proof, before the generation of the secret.
\item Once, they all get convinced with all the transactions, they will reveal their transaction IDs and the amount credited or debited from their account. Finally, they generate the secret $s_{B_k}$ for the $k^{th}$ block, where $s_{B_k} = \mathcal{E}(N_{trans}, N_{peop}, T_k, A_{deb}, A_{cred}) $ is a $m$-tuple such that first $t$ bits are message bits and remaining $m-t$ bits are parity bits, $N_{trans}$ is number of transactions happened in that particular time interval, $N_{peop}$ is number of people involved in the transactions, $T_k$ is same as defined above, and $A_{deb}$ and $A_{cred}$ represents the total amount debited and credited respectively (note that $A_{deb} = A_{cred}$), and $\mathcal{E}$ is the encryption function which maps $5$-tuple to $m$-tuple.
% comment: define the domain and codomain of \mathcal{E}
\item Dealers will then submit $s_{B_k}$ to the system and system will run our MSS and distribute the shares to random $m$ participants (active nodes) using Target-Anonymous Channel. Any $t$ nodes %(using the threshold of atleast $50\%$)
can compute the secret and then verify and validate the transactions within the given $\tau_{1}$ minutes.
\item Once, the verification and validation is done, a new block will be added to the chain using all the parameters required for block header and block generation.

\end{enumerate}
\end{example}

\subsection{Applications on various sectors}
We can effectively apply our scheme on different sectors such as national security, healthcare, supply chain management, decision making process of a company, elections, etc., where a few crucial and confidential information is required to be shared with a group of people in such a way that no adversary will get any information about it. A few of them are mentioned below:
\begin{enumerate}
\item \textit{National security} is a serious concern. Even a small attack or information leakage can have major consequences. Thus, we can use this scheme to protect the data. Also, authorities from different departments can communicate and take the decisions accordingly. For example, Nuclear Command Authority (NCA) of India, which is responsible for command, control and operational decisions regarding India's nuclear weapons programme, can interact with the Political Council headed by the Prime Minister of India and an Executive Council headed by the National Security Advisor, to take a decision regarding a nuclear test, in such a way that no outsider will get the information prior to the completion of the test.
\item If the board of directors of a \textit{company} takes a crucial decision which can affect the overall growth of the company, then shareholders can verify if the decision taken by the board will add to the future growth of the company or not and they can question it accordingly.
\item In \textit{healthcare} sector, patients can share their medical history (that includes medications, health issues, lab results etc.) with the hospital, termed as \textit{health information exchange} (HIE) and hospital can further forward it to specialists and other relevant departments within it . Also, it helps in storing the electronic health record (EHR) of the patient.
\item To apply our scheme on \textit{supply chain}, three basic entities: suppliers, enterprises, and market dealers can be considered. Enterprise can send their requirement and ask for the quotations from the suppliers in an encrypted form through blockchain platform. Similarly, enterprises can share their product information and quotation with market dealers.
\end{enumerate} \par

%%%%%%%%%%%%%%%%%%%%%%%%%%%%%%%%%%%%%%%
%%%%%%%%%%%%%%%%%%%%%%%%%%%%%%%%%%%%%%%
%%%%%%%%%%%%%%%%%%%%%%%%%%%%%%%%%%%%%%%
%%%%%%%%%%%%%%%%%%%%%%%%%%%%%%%%%%%%%%%

\section{Analysis}\label{section5}
\subsection{Analysis of Multisecret-sharing scheme}
Now we show here that our scheme is secure by proving that any $t-1$ or less participants can't retrieve the secret. If possible, we assume that $t-1$ participants come together to compute the secret \textbf{s}. For this they initially require to compute the polynomial $h(x)$. \par 
Without loss of generality, we assume $P_1, P_2, \dots , P_{t-1}$  are $t-1$ participants and $(a_i, h(a_i))$ are their respective shares. Also, $h(x)$ is a polynomial of degree $t-1$ with $t$ coefficients. Thus we have a system of $t-1$ linear equations in $t$ variables;
\begin{equation*}
\begin{split}
h(a_1) =& H_0 + H_1a_1 +\dots + H_{t-1}a_1^{t-1},\\
h(a_2) =& H_0 + H_1a_2 +\dots + H_{t-1}a_2^{t-1},\\
\vdots &\\
h(a_{t-1}) =& H_0 + H_1a_{t-1} +\dots + H_{t-1}a_{t-1}^{t-1}.
\end{split}
\end{equation*}
Therefore, 
\[
\underbrace{\begin{bmatrix}
1     &    a_1     & \dots     &       a_1^{t-1} \\
1       &   a_2     & \dots     &          a_2^{t-1}    \\
1       &   a_3     & \dots     &     a_3^{t-1}  \\
\vdots  &  \vdots  &   \ddots  &    \vdots \\   
1       &   a_{t-1}     & \dots     &   a_{t-1}^{t-1}  \\
            \end{bmatrix}
            }_{\mathbf{A}}
\underbrace{\begin{bmatrix}
H_0     \\
H_1     \\
H_2     \\
\vdots  \\
H_{t-1}     \\
            \end{bmatrix}
            }_{\mathbf{X}}
    =
\underbrace{\begin{bmatrix}
h(a_1)     \\
h(a_2)      \\
h(a_3)    \\
\vdots  \\
h(a_{t-1})     \\
            \end{bmatrix},
            }_{\mathbf{B}}
\]
where $A$ is a $(t-1) \times t $ matrix. Since, all $a_i's$ are distinct and it is a sub matrix of a Vandermonde matrix of size $t \times t$, implies rank of $A$ is $t-1$. For every $H_0 \in \mathbb{F}_p$ there exist unique $H_1, H_2, \dots , H_{t-1}$, which implies there are at least $p$ solutions. Thus, our scheme is secure against the attack made by any $t-1$ or less participants. Therefore, our scheme is perfect.

\iffalse \subsection{Efficiency}

To check the efficiency of our scheme, we require to compute information rate $\rho$ which is equal to the ratio of the length of
set of all possible secret and set of possible shares.
\begin{enumerate}
\item Information rate of secret sharing scheme $\rho_{s}=\frac{q-1}{q-1}=1$.
\item Information rate of multisecret-sharing scheme $\rho_{m}=\frac{•}{•}$
\end{enumerate}\fi

\subsection{Analysis of the scheme on Blockchain Network}
\begin{itemize}
	\item \textbf{Privacy}: Dealers first need to convince each other regarding their valid transactions using zero-knowledge proof and then secret will be generated using the encryption of the transaction details. Moreover, the honesty of the participants will be tested. Thus this scheme maintains privacy.
	
	\item \textbf{Integrity}: Data is stored after two-level verification (using our scheme) in blockchain network and once it is recorded in a block, it can't be removed. Also, each block is linked with the previous block hash and any change in the transaction will lead to the change in the hash value of all preceding blocks. Thus, data stored is immutable and permanent.
	
	\item \textbf{Security}:
	In our blockchain network, all nodes will be treated equally and new nodes can join only after signing smart contract and a proper verification by active nodes. Further, they require to prove their honesty before getting any information (shares) of the secret. 
Also each block is added to the blockchain network only after verifying and validating it by at least $t$ participants. Thus, it will provide security against double spending.	
Also, our scheme is secured against Finney attack, Race attack, 51\% attack, and Sybil attack.
\iffalse  DDoS attack (this attack can be prevented by ensuring that nodes have adequate storage, processing power and network bandwidth).\fi
	
\item \textbf{Scalability}:	Scalability in blockchain refers to the ability of the platform to expand as per the requirement and support the increasing load of transactions and nodes in the network.\par 
Performance of blockchain network is measured on the basis of average time taken by a transaction to validate. An increase in the number of nodes will lead to an increase in number of transactions, which will affect its performance. Each transaction require space to get stored in block. Moreover, blockchain is decentralized; thus each node is required enough space to store the data, which increases the storage and maintenance cost. Also every node must keep an updated record which will decrease the transmission speed.\par 
To resolve these issues, we have designed our algorithm in such a way that there will be only limited number of nodes. Moreover, secret sharing data can be deleted after the secret gets recovered and block formation process is done.\par 
Also to resolve the storage issue for every node, a few super computers can be installed which store the data in place of each node. It will also protect the network from single point of failure. However, nodes can be given access to that information. We can also use the method of sharding which involves splitting a blockchain into multiple pieces (called shards), and storing them at different places. It helps to manage the storage and cost problem with the increase in the number of transactions. \par 
Moreover, limited number of nodes and limited transactions will enhance the speed of transmission and compacting multiple transactions into an $m$ length secret will also reduce the storage requirement.\par 
Since each transaction holder has already convinced other dealers regarding the validity of the transaction, we have assumed if participants were not able to conclude (that the transaction is valid or not) within a pre-determined time interval, it would be considered as validated and automatically be added to the block. In this way, we can some how reduce the scalability issue in our MSS based blockchain network.
\end{itemize}

%%%%%%%%%%%%%%%%%%%%%%%%%%%%%%%%%%%%%%%
%%%%%%%%%%%%%%%%%%%%%%%%%%%%%%%%%%%%%%%
%%%%%%%%%%%%%%%%%%%%%%%%%%%%%%%%%%%%%%%
%%%%%%%%%%%%%%%%%%%%%%%%%%%%%%%%%%%%%%%

\section{Conclusion}\label{section6}
In this manuscript, we introduce $(t,m)$-threshold secret sharing scheme and multisecret-sharing scheme with two-level security based on Shamir's SSS using one-way function. Then we generalize the scheme to multi dealer (called as committee of dealers) to efficiently apply it to the blockchain network.

\end{document}